\documentclass[prb,aps,showpacs,twocolumn,unsortedaddress]{revtex4}
\usepackage{graphics,bm}
\usepackage{amssymb}
\usepackage{epsfig}
\usepackage{epsf}

\begin{document}

\title{Bias dependence of magnetic exchange interactions: application to interlayer exchange coupling in spin valves}
\author{Paul M. Haney$^{1}$}
\email{paul.haney@nist.gov}
\author{Christian Heiliger$^{1,2}$}%
\author{Mark D. Stiles$^{1}$}

 \affiliation{$^{1}$Center for Nanoscale Science and Technology, National Institute of Standards and Technology,
Gaithersburg, Maryland 20899-6202, USA }
\affiliation{
$^{2}$Maryland NanoCenter, University of Maryland, College Park,
Maryland 20742, USA }

\begin{abstract}
We study how a bias voltage changes magnetic exchange interactions.
We derive a general expression for magnetic exchange interactions
for systems coupled to reservoirs under a bias potential, and apply
it to spin valves. We find that for metallic systems, the interlayer
exchange coupling shows a weak, oscillatory dependence on the bias
potential. For tunneling systems, we find a quadratic dependence on
the bias potential, and derive an approximate expression for this
bias dependence for a toy model.  We give general conditions for
when the interlayer exchange coupling is a quadratic function of
bias potential.
\end{abstract}

 \maketitle

\section{Introduction}
Magnetic multilayers and spin valves exhibit a rich and extremely
useful array of magneto-electronic phenomena, including
magnetoresistance, interlayer exchange coupling, and spin transfer
torque.  Spin transfer torque is an effect in which the application
of a current can change the relative orientation of the layers'
magnetization. The presence of spin transfer torques are understood
as a result of nonequilibrium electron current flow, or more
precisely, a spatially varying spin current.  Spin currents describe
the flow of electron spin, and are tensor objects with two indices:
one labeling the direction of flow in real space, the other labeling
the vector component of spin. Interlayer exchange coupling is an
indirect exchange interaction between magnetic layers, like the
Ruderman-Kittel-Kasuya-Yosida (RKKY) interaction, with equilibrium
electrons in the nonmagnetic spacer playing the mediating role.
Interlayer exchange coupling is closely related to spin transfer
torques, as we discuss below. In this paper, we elucidate the
relationship between interlayer exchange coupling and spin transfer
torques, specifically how the interlayer exchange coupling depends
on the applied bias.

Interlayer exchange coupling gives rise to an energetic preference
for parallel or anti-parallel alignment of the layers'
magnetization, and is given approximately as $E=-J \hat M_1 \cdot
\hat M_2 $, where $\hat M_i$ are the directions of the
magnetizations. It shows RKKY-like characteristics, such as an
oscillatory dependence on spacer layer thickness, with a period
determined by the spacer material Fermi wave vector. This behavior
is well established \cite{reviews1,reviews2} by the close comparison
between the experimental \cite{parkin, unguris} and theoretical
results\cite{edwards,bruno,stiles}. The interlayer exchange coupling
can be computed using the general formalism for calculating magnetic
exchange interactions in bulk ferromagnets \cite{leich}. Slonczewski
has pointed out that interlayer exchange coupling can also be
understood as a consequence of conservation of total spin angular
momentum and the presence of spatially varying spin currents in
equilibrium (by equilibrium we mean no net charge current flow)
\cite{slonc89}.   In a spin valve, these equilibrium spin currents'
spin vector is perpendicular to the magnetization of both magnetic
layers, or out of the plane spanned by the two magnetizations. This
out-of-plane spin current results in an out-of-plane torque, denoted
by $\Gamma_\perp$, which is responsible for the interlayer exchange
coupling. Its magnitude is related to the interlayer exchange
coupling, and is given by $\Gamma_\perp = J |\hat M_1 \times \hat
M_2|$.

\begin{figure}[h!]
\begin{center}
\vskip 0.2 cm
\includegraphics[width=3.0in]{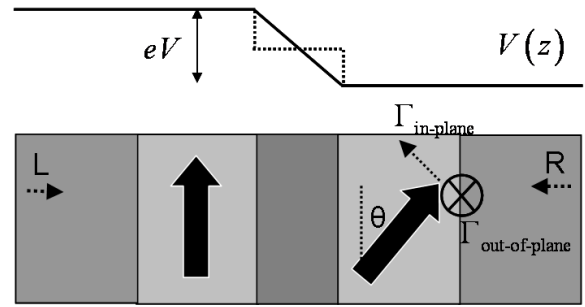}
\vskip 0.2 cm \caption{Spin valve showing the different components
of torques on a magnetic layer.  Bold arrows indicate the direction
of magnetization in the two ferromagnetic layers (light grey)
surrounded by nonmagnetic material (dark gray).  The assumed
potential drop is shown above the multilayer: the dark curve shows
the linear potential drop used for the numerical results, while the
dashed step-like function is the potential drop assumed in analytic
formulas.  Contributions to the torque are found from states
emanating from L and R lead separately. }\label{fig:multilayer}
\end{center}
\end{figure}

From the perspective of magnetic exchange interactions as arising
from the presence of spatially varying spin currents, it is a small
conceptual step to deduce the presence of spin transfer torque when
there is a nonzero net charge current.  Indeed, typical spin
transfer torques are the result of a net spin current with spin
vector in the plane spanned by the two magnetizations
\cite{slonc89,slonc,berger}, leading to torques in the plane,
denoted by $\Gamma_\parallel$. Frequently, the equilibrium torque is
referred to as the interlayer exchange coupling and the additional
torque due to current is referred to as the spin-transfer torque. We
find it conceptually useful to refer to the in-plane torque as the
spin transfer torque and the equilibrium plus the bias-dependent
out-of-plane torque as the interlayer exchange coupling. (Note
others have referred to the bias-dependent part of $\Gamma_\perp$ as
the ``field-like spin transfer term", as its form is the same as the
torque from an applied field.) The physical difference between
interlayer exchange coupling and spin transfer torque is then due
simply to the different spin vector components of the spin current
(see Fig. \ref{fig:multilayer}). We use the terms interlayer
exchange coupling and out-of-plane torque $\Gamma_\perp$
interchangeably.

The fact that $\Gamma_\perp$ and $\Gamma_\parallel$ originate from
different components of the spin current's spin vector imply further
qualitative differences between the torques. The most obvious
difference is that $\Gamma_\perp$ is present in equilibrium, while
$\Gamma_\parallel$ is only present in nonequilibrium systems. This
difference can be understood on general symmetry grounds
\cite{symmetryfootnote}. Another important difference is the
relative magnitude of $\Gamma_\perp$ and $\Gamma_\parallel$.  For
current densities attained in magnetization switching experiments
($10^{12} {\rm A/m^2}$), $\Gamma_\parallel \gg \Gamma_\perp$ for
metallic systems, which can be understood on general grounds and is
discussed further in Section (\ref{sec:results1}). Finally, there is
generally a qualitative difference in the dependence of the in-plane
and out-of-plane torques on the applied bias.  (We note that this
different bias-dependence of $\Gamma_\perp$ and $\Gamma_\parallel$
renders comparisons of their magnitude only partially meaningful.)
This difference in bias-dependence is the motivation for this work,
as there are inconsistent experimental results on the out-of-plane
torque bias-dependence.  For tunneling systems, a dependence of the
interlayer exchange coupling on the measured bias potential has been
reported as linear \cite{tulapurkar} or quadratic \cite{sankey},
while for metallic systems, experiments have shown that interlayer
exchange coupling is nearly independent of bias \cite{sankey}, while
others are interpreted as indicating a linear dependence with
substantial slope \cite{zimmler}.  Theoretical work has implied a
linear relation of the interlayer exchange coupling with bias for
metallic systems \cite{haney,heiliger2}, or posited a quadratic
dependence \cite{slonc05,theodonis}.  Ref. \onlinecite{xiao} proves
on symmetry grounds a quadratic dependence for symmetric systems
with semi-infinite ferromagnetic leads.

In light of these disparate results, this work considers more
carefully the bias dependence of interlayer exchange coupling (or
$\Gamma_\perp$). The question of how magnetic exchange interactions
are altered in the presence of current flow has been considered
before in Ref. \onlinecite{kozub}, which studied how the RKKY
interaction between two spins changes due to a current-carrying
electron distribution, and in Ref. \onlinecite{wingreen}, which
finds a bias-dependent exchange interaction proportional to the
ferromagnetic layer thickness. Heide proposed that effects now
understood to be the result of spin transfer torques were a
manifestation of current-altered exchange interactions \cite{heide}.
We add to these works in the hope of addressing recent experimental
questions, and to express bias-dependent exchange interactions in a
language familiar to previous studies of equilibrium exchange
interactions and interlayer exchange coupling.  Our general results
are valid for both multilayers and bulk ferromagnets in the regime
of ballistic transport.

We restrict our attention here to spin valves.  We find that the
bias dependence of interlayer exchange coupling can exhibit a wide
array of behaviors, so we focus here on some special cases in which
the physics is clear and which are relevant for realistic systems.
We find for metallic systems in the limit of weak magnetic
potentials and large spacer thickness that the interlayer exchange
coupling shows oscillatory dependence on bias; however for
experimentally attainable bias potentials, the change in interlayer
exchange coupling is negligible.  In the symmetric half-metallic
metal and tunneling limits, the interlayer exchange coupling depends
quadratically on the bias.  We derive approximate expressions for
the prefactor of this quadratic term and compare them to exact
numerical results.


\section{Method}

The approach we adopt is relevant for systems in the ballistic
limit.  In the spirit of Landauer's approach to transport in
mesoscopic systems, we suppose the system is connected  to two
reservoirs (to the left (L) and right (R)) in separate thermal
equilibria, with chemical potentials $\mu_{\rm L}$ and $\mu_{\rm
R}$. A bias voltage $eV$ across the system is represented by a
difference $eV$ between the reservoir chemical potentials.  States
in the system can be classified as emanating either from the L or R
lead (our formalism omits contributions from bound states with
energies within the bias window). We find expressions for the
interlayer exchange coupling due to states emanating from the L and
R lead separately. This permits us to find the interlayer exchange
coupling when L and R leads have different chemical potentials (that
is, when there is an applied bias).

We utilize a nonequilibrium Green's function approach, in which the
presence of the Left/Right lead is accounted for with a retarded
(advanced) self-energy $\Sigma^{r(a)}_{{\rm L/R}}(E)$, and
observables are expressed in terms of Green's functions
$G^{r(a)}(E)=\left(E-H-\Sigma^{r(a)}_{\rm L}(E)-\Sigma^{r(a)}_{\rm
R}(E)\right)^{-1}$.  Our approach generalizes the expressions of
Liechtenstein {\it et al.} \cite{leich} to find magnetic exchange
interactions for nonequilibrium systems.

When the spins of some subset of orbitals $\{i\}$ are rotated by an
angle $\theta$ from their collinear ground state orientation, there
will be exchange torques on these orbitals of the form
$J=J_1\sin(\theta) + J_3 \sin^3(\theta) + ...~$. Assuming $\mu_{\rm
L}>\mu_{\rm R}$, we state here the final result for exchange torque
coefficient $J_1$ (we defer the derivation, and more details
regarding nonequilibrium Green's functions, to appendix A):

\begin{eqnarray}
J_1 &=& \frac{1}{\pi}{\rm Im}  \int_{-\infty}^{\mu_{\rm R}} {\rm Tr}
\left[ \Delta \left(G^r_\uparrow - G^r_\downarrow \right) - \Delta
G^r_\uparrow \Delta G^r_\downarrow \right] dE \nonumber
\\ &&~~+{\rm Re} \int_{\mu_{\rm R}}^{\mu_{\rm R}+eV} {\rm Tr}
\left[\Delta\left(G^<_\uparrow- G^<_\downarrow \right) - \nonumber \right. \\
&& \left. ~~~~~~~~~~~~~~~~~~ \left(\Delta G^r_\uparrow \Delta
G^<_{L,\downarrow}  + \Delta G^<_{L,\uparrow} \Delta
G^a_\downarrow\right) \right]dE \label{eq:j}\nonumber\\,
\label{eq:JLR}
\end{eqnarray}
where $\Delta=H_\uparrow-H_\downarrow$ is the spin-dependent
Hamiltonian, $G^<_{L,\sigma} = \frac{i}{2\pi} G^r_\sigma
\left(\Sigma_{\rm L}^r-\Sigma_{\rm L}^a\right)_\sigma G^a_\sigma$,
and the traces are over orbitals $\{i\}$ that have rotated spins.
For ${\rm V}=0$, the limits on the 2nd integral are identical so
that its contribution vanishes. In this case Eq. (\ref{eq:JLR})
reduces to the well established expression for equilibrium exchange
interaction of Ref. \onlinecite{leich}.  For ${\rm V}\neq 0$, in
addition to contributions from the second term of Eq.
(\ref{eq:JLR}), the bias potential will also alter the Greens
functions.  Eq. (\ref{eq:JLR}) is applicable for both bulk
ferromagnets and magnetic multilayers. (In our previous ${\it
ab~initio}$ work on spin transfer torques in Co-Cu spin valves, we
presented only the contribution from the second term of Eq.
(\ref{eq:JLR}) \cite{haney,heiliger2}.)

As an application of Eq. (\ref{eq:JLR}), we consider a
one-dimensional, single band tight-binding model of two finite
ferromagnetic layers with $n_{{\rm F1}}, n_{{\rm F2}}$ sites,
separated by a nonmagnetic spacer with $n_{\rm S}$ sites, with
nonmagnetic leads. The model is parameterized by the spin-splitting
of the ferromagnetic (FM) layers (assumed equal) $\Delta$, a scalar
potential in the spacer layer $U$, the Fermi energy $E_{\rm F}$, and
the number of layer sites $n_{{\rm F1}}, n_{{\rm F2}}, n_{{\rm S}}$.
$n_{\rm S}$ determines the spacer layer width $D$: $D=a(n_{\rm
S}+1)$ for a lattice spacing $a$. Energies are presented in
dimensionless form, scaled by the tight-binding hopping parameter
$t$.  We focus here on some limiting cases in which the physics is
clear, and which are relevant for realistic systems. In all of our
numerical calculations we assume that the bias potential drop is
linear and occurs over the spacer layer (see Fig.
(\ref{fig:multilayer})), although in our limiting analytic results
we neglect this linear potential and assume a step-like potential.

\section{Results}

\subsection{Small confinement, large spacer}\label{sec:results1}
In the spirit of earlier work on interlayer exchange coupling, we
first consider weak spin-dependent potential in the ferromagnet
$V_{\uparrow,\downarrow}$, and large spacer thickness $D$. In this
limit, the one-dimensional, zero bias interlayer exchange coupling
takes the well-established form\cite{edwards,bruno,stiles}:
\begin{eqnarray}
J&=&\frac{\hbar v_{\rm F}}{2\pi D}{\rm
Im}\left[\left(r^{\uparrow}_{1}-r^{\downarrow}_{1} \right)
\left(r^{\uparrow}_{2}-r^{\downarrow}_{2}\right) e^{2ik(E_{\rm F})D}
\right], \label{eq:Jeq1}
\end{eqnarray}
where $r^{\uparrow(\downarrow)}_{1(2)}$ is the reflection off of the
spin- up (down) potential of layer 1 (2), and $k(E_{\rm F})$ is the
Fermi wave-vector of the spacer layer.  $v_{\rm F}$ is the Fermi
velocity, given by $v_{\rm F} = \partial E /\hbar \partial k$. (Note
that $r^{\uparrow(\downarrow)}_{1(2)}$ includes the thickness
dependence of the ferromagnetic layers.)  The interpretation of Eq.
(\ref{eq:Jeq1}) is that to lowest order in $r$, the two layers are
magnetically coupled together via itinerant states in the spacer
through spin-dependent reflection of spacer states on both layer 1
and 2, with the mediating state picking up a phase factor $2k(E_{\rm
F})D$ for the round trip between the layers. Eq. (\ref{eq:Jeq1}) has
been successful in describing the spacer material and thickness
dependence of interlayer exchange coupling in experiments
\cite{stiles, bruno, unguris}.

At zero bias, the interlayer exchange coupling on layers 1 and 2 is
necessarily equal and opposite, because there is no external source
of angular momentum. Therefore the choice of rotating layer 1 or 2
to find the interlayer exchange coupling is arbitrary. Under bias,
the interlayer exchange coupling on layers 1 and 2 can be different
because the reservoirs supply angular momentum when out of
equilibrium.  This implies that the interlayer exchange torque is
not the same on layers 1 and 2.  Here we consider the interlayer
exchange torque on layer 1. As described in the previous section, we
find the torque $J_{\rm L/R}$ due to electrons from the ${\rm L/R}$
lead separately. The calculation of $J_{{\rm L/R}}$ can be found in
Appendix B, here we give the result:
\begin{eqnarray}
J_{\rm L}&=&\frac{\hbar v_{{\rm F}}}{2\pi D} {\rm Im}\left[\left(r^{\uparrow}_{1}-r^{\downarrow}_{1} \right)\left(r^{\uparrow}_{2}-r^{\downarrow}_{2} \right) e^{2ik(E_{\rm F})D} \right], \nonumber\\
J_{\rm R}&=&0. \label{eq:JsmallR}
\end{eqnarray}

\begin{figure}[h!]
\begin{center}
\vskip 0.2 cm
\includegraphics[width=3.0in]{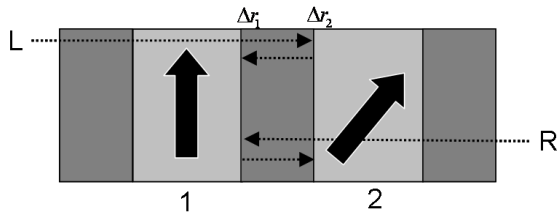}
\vskip 0.2 cm \caption{Paths for states emanating from L and R leads
for interlayer exchange coupling to lowest order in the
spin-dependent reflectivity $\Delta r$. }\label{fig:paths}
\end{center}
\end{figure}

We find that that all of the out-of-plane torque on layer 1 is due
to states emanating from L, while states from R contribute nothing.
This can be understood in terms of paths of states from L or R
considered to second order in the reflectivity in $r$.  Fig.
(\ref{fig:paths}) shows the relevant path for states from the L and
R leads.  A state from L transmits through 1, reflects off 2 (so
that it now ``carries" information about 2), comes back and reflects
off 1, effectively ``communicating" that information to 1, thereby
coupling the two layers.  For states from R, the two-reflection path
transmits through 2 (so is initially oblivious of it), reflects off
1, comes back and reflects off 2. A final reflection off 1 would
communicate spin information from 2, but this would be a third order
process. There is therefore no second order contribution to the
interlayer exchange coupling on layer 1 with this path.
There are, of course, contributions to both $J_{\rm L}$ and $J_{\rm
R}$ that are higher order in the reflectivities.  However, in the
limit that the reflectivities are much less than one, the additional
contributions are negligible.

\begin{figure}[h]
\begin{center}
\vskip 0.2 cm
\includegraphics[width=3.25in]{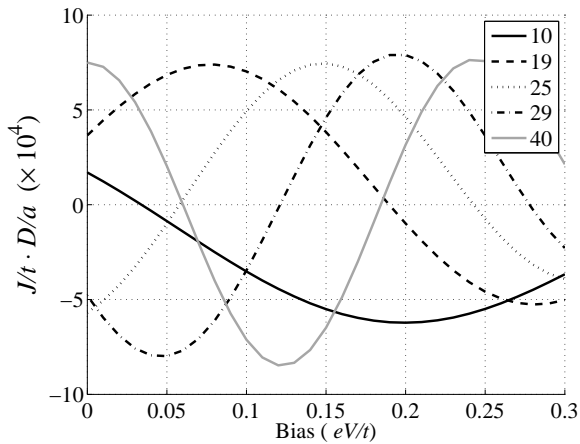}
\vskip 0.2 cm \caption{$\left(J/t\cdot~D/a \right)$ vs. bias
potential for several spacer layer thicknesses.  $n_{{\rm
F1}}=n_{{\rm F2}}=7,~ \Delta=0.1t,~ U=0, ~ E_{\rm F}=-0.78t$ for our
1-d model. Legend shows different spacer widths in units of the
lattice spacing $a$.}\label{fig:jdata1}
\end{center}
\end{figure}

We assume that the dependence of $J$ on $E_F$ is dominated by the
exponential factor. (That is, we assume $D \left(\partial k /
\partial E\right)$ is much larger than $\partial r / \partial E$ and $~(\hbar/JD)(\partial v_{\rm F} /
\partial E)$, which is valid in the large $D$ limit). The bias dependence
can then easily be found:
\begin{eqnarray}
J(V) = \frac{\hbar v_{{\rm F}}}{2\pi D}{\rm
Im}\left[\left(r^{\uparrow}_{1}-r^{\downarrow}_{1} \right)
\left(r^{\uparrow}_{2}-r^{\downarrow}_{2}\right) e^{2ik\left(E_{{\rm
F}}+eV_{{\rm L}} \right)D} \right],\label{eq:JV1}
\end{eqnarray}
where $eV_{{\rm L}}$ is the bias potential difference between left
lead and spacer layer, and the wave vector $k$ is a function of the
argument $E_{\rm F}+eV_{{\rm L}}$. The interlayer exchange coupling
is an oscillatory function of the bias $V$, with period $2\pi \hbar
v_{{\rm F}} / D$.  Fig (\ref{fig:jdata1}) shows numerical evaluation
of Eq. (\ref{eq:JLR}) for the 1-dimensional tight binding model, and
we find excellent agreement with the approximate $J(V)$ of Eq.
(\ref{eq:JV1}) in the large $D$ regime. We note that the {\it phase}
of the oscillation is unconstrained, and is set by model parameters,
most experimentally relevantly by $D$.

The success of Eq. (\ref{eq:Jeq1}) in describing experimental data
encourages us to believe that Eq. (\ref{eq:JV1}) is also
experimentally relevant for metallic systems.  Making a stationary
phase approximation and considering only extremal wave vectors of
the Fermi surface, the maximal possible slope of $J(V)$ near 0 bias
is:
\begin{eqnarray}
{\frac{dJ}{dV}}_{\rm max} = |J_0| \frac{De}{\hbar v_{{\rm F}}^*}~,
\label{eq:slope}
\end{eqnarray}
where $e$ is the electron charge, and $|J_0|$ is the maximum
equilibrium value (in the 1-d model, $|J_0|=\hbar v_{\rm F}|\Delta
r_1\Delta r_2|/2 \pi D$), which in this approximation is also
related to the value of the coupling at the extremal wave vector,
and includes prefactors such as the curvature of the Fermi surface
at the extremal point, (see Refs. \onlinecite{bruno,stiles} for
example), and $v_{{\rm F}}^*$ is the Fermi velocity of the extremal
wave vector. In the case of multiple extremal wave vectors, Eq.
(\ref{eq:slope}) should be a sum over all of them.  For a Cu spacer
with $D=10 ~{\rm nm}$, the change in $J$ is $|J_0|0.971~{\rm V}$.
Assuming the maximum voltage on a metallic multilayer to be on the
order of $10^{-4}~{\rm V}$, this leads to a maximum change on the
order of $10^{-4}|J_0|$ (typical values for $J_0$ are $0.24~{\rm
mJ}/{\rm m}^2,$ \cite{bloemen}). We therefore conclude that in
metallic systems, the bias-induced change in interlayer exchange
coupling is negligible and not measurable.

The oscillatory dependence of the exchange coupling (or out-of-plane
torque) on bias may be surprising in light of the general argument
for quadratic dependence in symmetric junctions given in Ref.
\onlinecite{xiao}. It is important to note that this argument
assumes semi-infinite ferromagnetic layers, and therefore does not
apply to the system under consideration here.


It is instructive at this point to contrast the physics of
current-altered interlayer exchange coupling and spin transfer
torques. We emphasize that the interlayer exchange coupling
properties are dominated by spacer state properties (such as $D$,
$k_{\rm F}$, and the curvature of the Fermi surface around extremal
points); in this limit, a bias changes interlayer exchange coupling
to the extent it changes these spacer state properties.  The
sensitivity of interlayer exchange coupling to spacer state
properties can be motivated by considering the interlayer exchange
coupling as resulting from out-of-plane spin currents.  For layers 1
and 2 in the $\hat x$ and $\hat z$ directions respectively, there is
no obvious natural preference for the out-of-plane spin current to
be in the positive or negative $\hat y$ direction. Indeed, the
values of out-of-plane spin current for different states are
distributed around 0, and mostly cancel. Only around extremal points
is there a net contribution (this is why expressions for interlayer
exchange coupling involve quantities evaluated only at $E_{\rm F}$
and at extremal wave vectors), and the final value is in some sense
accidental.  On the other hand, the in-plane torque acting on layer
2 has a strong preference to point in the $x$ direction.  As for the
out-of-plane torque, the specific values of in-plane torque from
different states will vary, but they are distributed now about a
nonzero value.  For typical metallic systems such as Co-Cu at a
current density of $J=10^8 {\rm A/cm^2}$, the magnitude of the
out-of-plane torque is $5~\%$ that of the in-plane, reflecting this
qualitative difference in the spin current vector. This point has
been made by Bauer {\it et al.} in the language of the mixing
conductance $G^{\uparrow\downarrow}$. There the effect is manifest
in a small imaginary part of
$G^{\uparrow\downarrow}$\cite{zwierzycki}.

\subsection{Half-metal}

In this section we take the limit of a metallic junction with
half-metallic FM layers.  We choose this system because it most
easily demonstrates the opposite limit of the small confinement
case.  We find that in the half-metallic case, electrons from the L
and R leads contribute equally to the interlayer exchange coupling
on each layer (whereas in the small confinement case, electrons from
only one lead contributed to the interlayer exchange coupling). Here
we assume the potential for majority channel is flat throughout the
magnetic and nonmagnetic layers, and the minority potential is
infinite in the ferromagnet.  The out-of-plane torque coefficient
(near antiparallel alignment, or $\theta \approx \pi$) then takes
the simple form:
\begin{eqnarray}
J_{{\rm L,R}} = \frac{1}{2\pi}\int_{-\infty}^{\mu_{{\rm L,R}}}
\sin\left[2k\left(E\right)D\right] dE~,
\end{eqnarray}
This expression can be found by computing the out-of-plane spin
current in the spacer layer. Integrating over all energies leads to
the exact result for the interlayer exchange coupling:
\begin{eqnarray}
J_{{\rm L,R}} &=& \frac{2 v_{\rm F} \hbar (D/a^2)\cos(2 k_F D) +
E_{\rm F}\sin(2 k_F D)}{4(D/a)^2-1}.
\end{eqnarray}

The bias dependence is given by simply shifting the energies of L
and R electrons by $+V/2$ and $-V/2$, respectively, leading to:
\begin{eqnarray}
J(V)&=&J_{\rm L}\left(E_{{\rm F}}+\frac{V}{2}\right) +
J_{\rm R}\left(E_{{\rm F}}-\frac{V}{2}\right) \nonumber \\
&\approx& J(0) + \frac{1}{2}\frac{\partial^2 J\left(E_{\rm
F}\right)}{\partial E^2} \left(\frac{V}{2}\right)^2 ~.\label{eq:v2}
\end{eqnarray}

\begin{figure}[h]
\begin{center}
\vskip 0.2 cm
\includegraphics[width=3.25in]{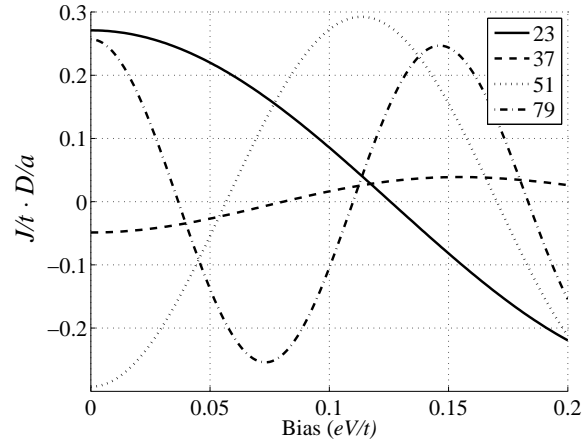}
\vskip 0.2 cm \caption{$\left(J/t\cdot~D/a \right)$ vs. bias
potential for several spacer layer thicknesses. $n_{{\rm
F1}}=n_{{\rm F2}}=7,~ E_{\rm F}=-0.78t$. FM are half metals and
results are for $\theta$ near $\pi$.
 Legend shows different spacer widths in units of the lattice spacing $a$.}\label{fig:jdata2}
\end{center}
\end{figure}

Fig. \ref{fig:jdata2} shows numeric results of the bias dependent
interlayer exchange coupling for a number of spacer widths.  We find
these results correspond well with Eq. (\ref{eq:v2}).  The curves
for interlayer exchange coupling are again periodic in $V$, with the
same period $2\pi \hbar v_{{\rm F}} / D$ as the previous small $r$
case. In contrast to the small $r$ case, the phase of the
oscillations are constrained so that all curves are $\cos$-like,
which leads to a quadratic dependence of the exchange torque at
small $V$.

\begin{figure}[h]
\begin{center}
\vskip 0.2 cm
\includegraphics[width=3.in]{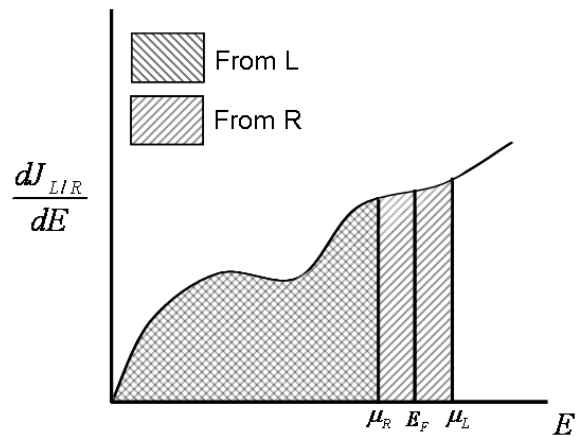}
\vskip 0.2 cm \caption{Schematic of how bias potential changes the
contributions from the L and R leads to the total interlayer
exchange coupling. At zero bias both L and R are contribute equally
up to $E_{\rm F}$, while a symmetric bias drop of $eV$ opens a
window from $E_{\rm F}-eV/2$ to $E_{\rm F}+eV/2$ with only one lead
contributing. This leads to a quadratic dependence of the coupling
on the bias. }\label{fig:JvsEatbias}
\end{center}
\end{figure}

The half-metallic system is emblematic of the general fact that if
states from the L and R leads contribute equally to the interlayer
exchange coupling, and if there is a symmetric bias potential drop,
then the bias dependence of the out-of-plane torque must be
quadratic.  Mathematically this quadratic dependence is seen
immediately from Eq. (\ref{eq:v2}).  A physical description is that
when a bias $eV$ is applied, we {\it gain} contributions from L
states at energies above the original Fermi energy (from $E_{\rm F}$
to $E_{\rm F}+\frac{eV}{2}$), while we {\it lose} contributions from
a {\it lack} of R states at energies below the original Fermi energy
(from $E_{\rm F}-\frac{eV}{2}$ to $E_{\rm F}$) (see Fig.
(\ref{fig:JvsEatbias})).
This gain and loss cancel each other to linear order in $eV$, and
only a quadratic dependence remains \cite{ralphreview}.  It should
be noted that in making a constant shift in energies of the left and
right leads relative to the spacer layer, there are also changes in
the contribution from the bottom of the bands.  However near the
bottom of the band, the contributions to $\Gamma_\perp$ oscillate
rapidly and don't contribute (due to a large $\frac{\partial
k}{\partial E}$), so that shifts in this region are unimportant. The
quadratic dependence of the out-of-plane torque is also discussed in
Refs. \onlinecite{xiao,theodonis} for the case of semi-infinite
ferromagnetic layers.

\subsection{Tunneling}

We next consider the case of a tunneling barrier with half-metallic
ferromagnetic layers.  Such a model system is relevant for the
states at the Brouillon zone center of Fe-MgO-Fe.  These states make
the dominant contribution to the transport \cite{butler}, spin
torques \cite{heiliger}, and interlayer exchange coupling
\cite{mgOinterlayer exchange coupling}. An exact expression for the
out-of-plane torque is given in Ref. \onlinecite{slonc89} for a
trilayer (FM leads separated by a spacer); for our 5-layer geometry,
$\Gamma_\perp$ is generally a complicated function of model
parameters.  We find that in the symmetric system ($n_{{\rm F1}} =
n_{{\rm F2}}$), electrons from both L and R leads contribute equally
to the interlayer exchange coupling, so that there is a quadratic
dependence of the interlayer exchange coupling on bias voltage, as
discussed in the previous section. We make a simplifying ansatz here
for $J_{{\rm L/R}}$ and compare its prediction to the $V^2$
prefactor with the exact numerical results. We find that our
approximate, analytic form shows qualitative agreement with the
numerics.

We suppose that the energy dependence of the out-of-plane torque is
dominated by the tunneling exponential:
\begin{eqnarray}
J_{{\rm L,R}} =  \int_{-\infty}^{\mu_{{\rm L,R}}} A
e^{-2\kappa\left(E\right) D} dE~,
 \label{eq:gammaE}
\end{eqnarray}
where $\kappa$ is the decay constant (or imaginary wave-vector) in
the spacer.  In our model, $\kappa a =  {\rm
Im}\left[\cos^{-1}\left(\left(E_{{\rm
F}}-U\right)/2t\right)\right]$.  $A$ is a complicated function of
model parameters, and we have found that compared to the exponential
factor, $A$ depends weakly on energy; we therefore omit its energy
dependence going forward.  To find the interlayer exchange coupling,
we assume energies near the Fermi level make the dominant
contribution to the total, leading to:
\begin{eqnarray}
J_{{\rm L,R}}= \frac{At\sinh\left(\kappa_{\rm F} a\right)}{D'}
e^{-2\kappa_{\rm F} D}~.
\end{eqnarray}
where $D'=D/a$.  The bias-dependent interlayer exchange coupling can
be written in the form:
\begin{eqnarray}
J(V) = J(0)\left(1+BV^2 \right).
\end{eqnarray}
A straightforward calculation leads to :
\begin{eqnarray}
B &=& \frac{2D'^2-D'\coth(\kappa a)-\frac{1}{2}~{\rm sech}^2(\kappa
a)}{8t^2\sinh^2(\kappa a)}~ .\label{eq:B}
\end{eqnarray}

\begin{figure}[h]
\begin{center}
\vskip 0.2 cm
\includegraphics[width=3.25in]{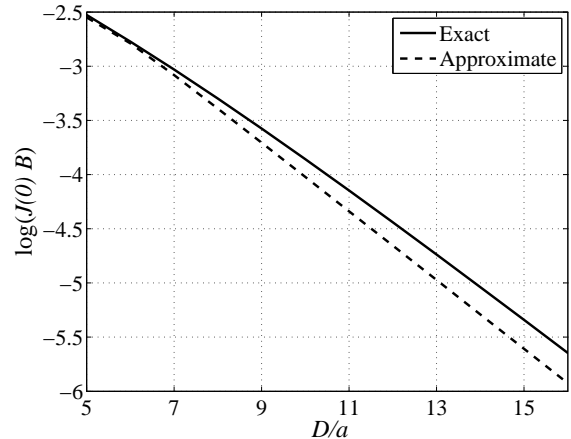}
\vskip 0.2 cm \caption{Logarithm of quadratic bias-dependence
prefactor versus spacer thickness $D$.  Parameters are $U=1.75t,
E_{{\rm F}}=-0.4t, n_{{\rm F1}}=n_{{\rm F2}}=7$.}\label{fig:Bvsn}
\end{center}
\end{figure}

\begin{figure}[h]
\begin{center}
\vskip 0.2 cm
\includegraphics[width=3.25in]{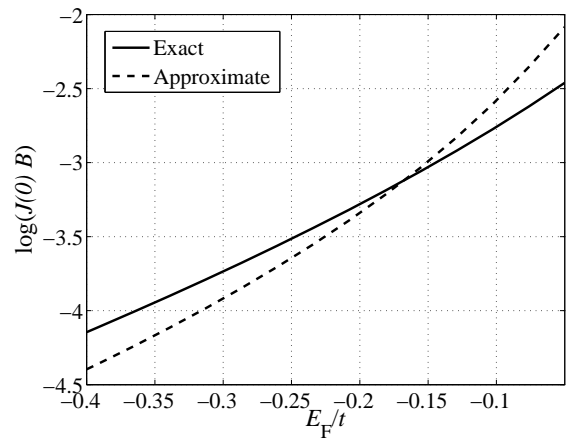}
\vskip 0.2 cm \caption{Logarithm of quadratic bias-dependence
prefactor versus Fermi energy.  Parameters are $U=1.75t, D/a=7,
n_{{\rm F1}}=n_{{\rm F2}}=7$.}\label{fig:Bvsef}
\end{center}
\end{figure}

In Fig. \ref{fig:Bvsn}, we plot the logarithm of the $V^2$ prefactor
for both exact numerical results and for the approximation given
above, as a function of $D/a$.  The agreement is good, given the
fact that we neglect the deformation of the barrier due to the
external bias potential in our approximation, and the qualitative
trend is correct.  Fig. \ref{fig:Bvsef} shows the prefactor as a
function of Fermi level, and again we find qualitative agreement.
The prefactor is exponentially sensitive to system parameters, due
to the fact that it is proportional to $J(0)$; the dependence of $B$
on system parameters is much weaker.
For all parameters we've considered, we find that the zero bias
exchange coupling between layers is antiferromagnetic in the
tunneling regime, and that the application of a bias increases this
antiferromagnetic coupling (i.e., the sign of the $V^2$ prefactor is
the same as the sign of $J(0)$).

Recent experiments on MgO indicate that at bias potentials on the
order of $0.5 ~{\rm V}$, the in-plane and out-of-plane torques are
of similar magnitude, in stark contrast to metallic systems in which
$\Gamma_\parallel \gg \Gamma_\perp$.  The origin of this difference
is the reduced area in $k$-space for the in-plane torque in Fe-MgO.
In metals all of the states in the Brouillon zone contribute to the
in-plane torque and only states near extremal points of the Fermi
surface contribute to the out-of-plane torque, while in Fe-MgO only
states near the $\Gamma$ contribution to both in-plane and
out-of-plane torques.

\section{Conclusion}

In this work, we have derived a general expression for magnetic
exchange interactions that is valid in nonequilibrium systems in the
ballistic regime.  We applied this formula to magnetic spin valves
to study how the interlayer exchange coupling depends on an external
bias voltage.  We find that for metallic systems in the small
confinement limit, the interlayer exchange coupling is an
oscillatory function of the bias, but that the period of the
oscillations is so long that seeing a change in interlayer exchange
coupling is not experimentally feasible.  For tunneling systems, and
for any system in which electrons emanating from the L and R leads
contribute equally to the interlayer exchange coupling and there is
a symmetric bias potential drop, we find that the interlayer
exchange coupling depends quadratically on the bias voltage, which
is in agreement with recent experiments.

We emphasize that the results we obtain are for systems in the
ballistic regime.  The behavior of $\Gamma_\perp(V)$ for systems in
the diffusive regime is generally different.  In particular, a
nonzero $\Gamma_\perp$ at zero bias is a phase coherent effect;
diffusive systems have a vanishing equilibrium $\Gamma_\perp$.  The
oscillatory $V$-dependence described for metallic systems is derived
from modulating the wave-vector of coherent states, and will not be
seen for diffusive systems.  Diffusive systems under bias generally
have a nonzero $\Gamma_\perp$.  For symmetric systems, this
dependence is quadratic and is therefore omitted from purely linear
theories such as the spin circuit theory \cite{brataas}.  For
asymmetric structures, there is a linear contribution to
$\Gamma_\perp(V)$, which is described by linear, diffusive theories,
where for example in the spin circuit theory it is related to the
imaginary part of the mixing conductance

There are other systems in which bias-altered exchange interactions
can have important effects.  It is possible, for example, that in
single ferromagnetic layers under bias, intralayer exchange
interactions can be affected by this bias, which would be manifest
in a change of spin wave dispersion or Curie temperature.  We are
currently investigating these types of systems.

\appendix* \section{Expansion of nonequilibrium density matrix in magnetic rotations}

Here we describe in more detail the calculation of exchange
interactions in systems under bias. Our approach is based on the
nonequilibrium Green's function method. In this method, we assume
there is some central region, described by a Hamiltonian $H_{\rm
C}$, embedded between two semi-infinite reservoirs or leads.  The
left and right leads are described by the semi-infinite Hamiltonians
$H_{\rm L}$ and $H_{\rm R}$ respectively.  We denote the coupling
between the central region and the left/right leads by
$\tau_L$/$\tau_R$. The retarded Green's function for the isolated
central region is simply $G_{\rm C}^r(E) = \left(E+i\eta-H_{\rm
C}\right)^{-1}$.  The presence of the left lead is accounted for
with self-energy $\Sigma_{\rm L}(E) = \tau_{\rm L} g_{\rm L}^{r0}
\tau_{\rm L}^\dagger$, where $g_{\rm L}^{r0}=\left(E+i\eta - H_{\rm
L}\right)^{-1}_{0,0}$ is the so-called surface Green's function.
This is the projection of the full semi-infinite lead Green's
function onto the subset of lead sites which couple to the central
region (for the one-dimensional model we consider, this subset is
denoted as the $({0,0})$ element of the full left lead Green's
function). The self-energy for the right lead is defined similarly.
The total retarded Green's function for the center region plus leads
is then $G^r(E) = \left(E + i\eta- H_{\rm C} - \Sigma_{\rm L}^r -
\Sigma_{\rm R}^r \right)^{-1}$.  In the following we combine left
and right self energies: $\Sigma^{r,a} \equiv \Sigma_{\rm L}^{r,a} +
\Sigma_{\rm R}^{r,a}$.

The so called ``lesser" Green's function is given by the well-known
Keldysh equation : $G^<=\frac{1}{2\pi}G^r\Sigma^<
G^a$\cite{haug,datta}, where $\Sigma^<_{\rm
L/R}=i\left(\Sigma^r_{\rm L/R}-\Sigma^a_{\rm L/R}\right)$. The
contribution to the density matrix from states emanating from the
L/R lead is:
\begin{eqnarray}
\rho_{\rm L/R} &=& \int_{-\infty}^{E_{\rm F}} G_{\rm L/R}^<(E)~ dE
\\&=& \frac{1}{2\pi}\int_{-\infty}^{E_{\rm F}}G^r \Sigma^<_{\rm L/R} G^a
dE. \label{eq:a1}
\end{eqnarray}
We consider how the density matrix changes upon the addition of some
perturbation $H'$ to the Hamiltonian.  To this end we consider how
the integrand of Eq. (\ref{eq:a1}) changes, and omit the energy
integral for clarity.  The integrand is given more explicitly as:
\begin{eqnarray}
G^<=\frac{1}{2\pi}\left( \frac{1}{E-H_{\rm C}-\Sigma^r}\right)
\Sigma^<\left(\frac{1}{E-H_{\rm C}-\Sigma^a}\right).
\end{eqnarray}
We restrict our attention to perturbations $H'$ which are only
present in the central region; $H'$ leads to a new Green's function
$G^<_1$, given by:
\begin{eqnarray}
G^<_1 &=&\frac{1}{2\pi}\left(\frac{1}{E-H_{\rm
C}-H'-\Sigma^r}\right) \Sigma^<\left(
\frac{1}{E-H_{\rm C}-H'-\Sigma^a}\right). \nonumber \\
\end{eqnarray}
To first order in $H'$, $G^<_1$ is:
\begin{eqnarray}
G^<_1&\approx &\frac{1}{2\pi}~G^r(1+H' G^r) \Sigma^< G^a (1+ H' G^a)  \nonumber \\
&\approx& G^< + G^r H' G^< + G^< H' G^a .\label{eq:expansion}
\end{eqnarray}

All of these objects are represented in a real-space basis, and we
consider perturbations about a collinear ground state.  The
components of the unperturbed Green's functions are then labeled
with site index $i,j$, and are diagonal in spin space:
\begin{eqnarray}
G_{i,j} =   \left(            \begin{array}{cc}
                  G_{\uparrow,i,j} & 0 \\
                  0 & G_{\downarrow,i,j} \\
                \end{array} \right).
\end{eqnarray}
In all $2\times 2$ matrices, we indicate explicitly site labels with
indices and also spin label. As before, we define $\Delta_{i,j} =
\delta_{i,j}\frac{1}{2}\left(H_\uparrow-H_\downarrow \right)_{i,j}$.
We consider tilting some portion (sites $\{i'\}$) of the
magnetization by a small angle $\theta$, so that $H'$ is given by:
\begin{eqnarray}
H'_{i,i} = \sum_{i\in \{i'\}} \frac{\Delta_{i,i}}{2} \left(
                \begin{array}{cc}
                  \cos(\theta)-1 & \sin(\theta) \\
                  \sin(\theta) & 1-\cos(\theta) \\
                \end{array}
              \right)~. \label{eq:perturbation}
\end{eqnarray}
There can in principle be spin-dependent terms that are off-diagonal
in site index, but these are usually much smaller than site-diagonal
terms and omitted here.  Inserting Eq. (\ref{eq:perturbation}) into
Eq. (\ref{eq:expansion}) yields:
\begin{eqnarray}
&&G^<_1 - G^<= \nonumber\\
  &&~~       \frac{1}{2}\left[
           \begin{array}{cc}
             G^r_\uparrow(\cos(\theta)-1)\Delta'G^<_\uparrow & G^r_\uparrow \sin(\theta)\Delta' G^<_\downarrow \\
             G^r_\downarrow \sin(\theta)\Delta' G^<_\uparrow  & G^r_\downarrow(1-\cos(\theta))\Delta'G^<_\downarrow \\
           \end{array}
         \right]+{\rm h.c.} \nonumber\\
\end{eqnarray}
Where for notational clarity we've introduced matrices $\Delta'$,
where $\Delta'_{i,i} = \Delta_{i,i}$ if $i \in \{i'\}$, and 0
otherwise.  In the above h.c. stands for ``Hermitian conjugate".
 From the perturbed density matrix we find the ensuing spin densities
(to first order in $\theta$):
\begin{eqnarray}
m_{i,j}^x &=& 2 \theta~{\rm Re} \left[ (G^r_\uparrow \Delta'
G^<_\downarrow + G^r_\downarrow \Delta' G^<_\uparrow)\right]_{i,j}  \nonumber \\
m_{i,j}^y &=& 2 \theta~{\rm Im} \left[ (G^r_\uparrow \Delta'
G^<_\downarrow - G^r_\downarrow \Delta' G^<_\uparrow)\right]_{i,j} \nonumber \\
m_{i,j}^z &=& \left(G^<_\uparrow - G^<_\downarrow \right)_{i,j} +
O\left(\theta^2\right) \label{eq:mexpansion}~.
\end{eqnarray}

As discussed in the introduction, the exchange torques in a
multilayer can be found by computing the out-of-plane torque. These
torques result from the misalignment between the spin-dependent
Hamiltonian and the spin density.  To find this misalignment, we
need the vector components of the spin-dependent perturbed
Hamiltonian $H+H'$, given as: $\Delta_\alpha = {\rm Tr}
\left[\left(H+H'\right)\sigma_\alpha\right], ~ \{\alpha=x,y,z\}$,
where $\sigma_\alpha$ are Pauli spin matrices.  If the layers'
magnetizations span the $x-z$ plane, then the exchange torque is:
\begin{eqnarray}
\Gamma_y = \Delta_x m_z  - \Delta_z m_x~.
\end{eqnarray}
Expanding to first order in $\theta$ leads to the following
expression for the coupling due to states from the L/R leads:
\begin{eqnarray}
J_{{\rm L/R}} &=& {\rm Re}~ \int_{-\infty}^{\mu_{\rm L/R}} {\rm Tr}
\left[ \Delta' \left(G^<_{\uparrow
\rm L/R}-G^<_{\downarrow \rm L/R}\right) - \nonumber \right. \\
&&\left. \left(\Delta' G^{r}_{\uparrow} \Delta' G^<_{\downarrow \rm
L/R} + \Delta' G^{r}_{\downarrow} \Delta' G^<_{\uparrow \rm L/R}
\right) \right] dE ~. \nonumber \\ \label{eq:jneq}
\end{eqnarray}

In equilibrium, the total lesser Green's function $G^<=G^<_L+G^<_R$
can be written in terms of the retarded Green's function alone:
$G^<=\frac{i}{2\pi}\left(G^r - G^a\right)$, in which case the
expression for the total exchange torque becomes:
\begin{eqnarray}
J=\frac{1}{\pi}{\rm Im}  \int_{-\infty}^{E_{\rm F}} {\rm Tr} \left[
\Delta' \left(G^r_\uparrow - G^r_\downarrow \right) - \Delta'
G^r_\uparrow \Delta' G^r_\downarrow \right] dE~. \nonumber
\\\label{eq:Jeq}
\end{eqnarray}
Combining Eq. (\ref{eq:Jeq}) for energies in which both leads are
occupied, and Eq. (\ref{eq:jneq}) for energies in which only one
lead is occupied yields Eq. (\ref{eq:JLR}).

We should note that in using the {\it non-self-consistent} spin
density from a system rotated slightly out of collinearity to find
torques, we are implicitly making use of the magnetic force theorem.
In addition, we utilize the same assumptions as Ref.
\onlinecite{leich}, namely that the orientation of spin magnetic
moments on each lattice site is a well-defined quantity, and that
these spin orientation degrees of freedom are ``slow" with respect
to the electronic degrees of freedom (adiabatic approximation).  The
validity of these approximations have been considered in detail in
previous works \cite{antropov}.

We make some final comments to compare our result for exchange
interactions with previous equilibrium expressions for exchange
interactions. First, in similar spirit to Ref. \onlinecite{leich},
we can attempt to define the pairwise interaction of two spins
$J_{i,j}$ by finding the exchange torque present on atom $i$ when
$i$ and $j$ are both rotated, and subtracting off the torques
present on $i$ when $i$ and $j$ are rotated separately.  Under
nonequilibrium conditions however, the following sum rule is
violated:
\begin{eqnarray}
J_i \neq \sum_{j\neq i} J_{ij}.
\end{eqnarray}
implying that a pairwise interaction is not properly defined in
nonequilibrium conditions.  This is again due to the fact that there
is a net flux of angular momentum into the system from the
current-carrying electrons.

Finally we note that the approach taken in Ref. \onlinecite{leich}
to finding exchange interactions relies on calculating the change in
single particle energy $\Delta E$ upon rotating a portion of the
magnetism away from collinearity.  It can be shown that in
equilibrium, this change in energy can be expressed in the language
of Greens functions as:
\begin{eqnarray}
\Delta E = \frac{1}{2}{\rm Tr}\left[H'\left(\rho' +
\rho\right)\right]~.
\end{eqnarray}
where $\rho'$ is the density matrix corresponding to the perturbed
lesser Green's function $G^<_1$.  We have found that this expression
works in nonequilibrium conditions as well, and gives identical
results to the exchange torques found using spin densities.

\subsection{In-plane torques}

With the expansion of the nonequilibrium density matrix given by
Eqn. (\ref{eq:expansion}), it is a simple matter to find the spin
transfer torque, which we do here for completeness.  As in Ref.
\onlinecite{haney}, we find the out-of-plane spin density to
determine the spin transfer torque present on a subsystem of atoms
$\{j\}$, when atoms $\{i'\}$ are rotated by a small angle $\theta$.
Assuming the two layers span the $x-z$ plane, the $y$-component of
the spin density is called for: the out-of-plane spin density in a
spin valve determines the spin torques.  The spin transfer torque is
thus $\Gamma^{IP}= |\Delta| m_y$.  As before, we use the expansion
of $\vec m$ given by Eq. (\ref{eq:mexpansion}) to obtain:
\begin{eqnarray}
\frac{\Gamma^{IP}}{\sin(\theta)}&=&\frac{\mu_B}{\hbar} ~{\rm Im}
\int_{\mu_{\rm R}}^{\mu_{\rm R}+eV}~{\rm Tr} \left[ \Delta'
G^r_\uparrow
\Delta'' G^<_\downarrow - \right.  \nonumber  \\
&&~~~~~~~~~~~~~~~~\left. \Delta' G^r_\downarrow \Delta''
G^<_\uparrow\right] dE~. \label{eq:stt}
\end{eqnarray}

Where the trace is over orbitals of the subsystem $\{j\}$ in
question, $\Delta'$ and $\Delta''$ are the spin-dependent
Hamiltonian for orbital sets $\{j\}$ and $\{i\}$, respectively.  Eq.
(\ref{eq:stt}) has the utility of expressing in plane torques in
terms of collinear objects, simplifying the numerical task of
calculating these torques from first principles relative to
calculating full noncollinear spin configurations, as in Refs.
\onlinecite{haney} and \onlinecite{heiliger2}.

\subsection{Bias dependence of interlayer exchange coupling}

Here we take the limit of large spacer thickness and small
reflectivities, and reformulate $J$ in the language of Refs.
\onlinecite{stiles} and \onlinecite{bruno}.  We consider first the
equilibrium case as a precursor to the nonequilibrium case. We can
rewrite the Green's functions in terms of free space Green's
functions $G_0$:
\begin{eqnarray}
G^r &=& G_0^r + G_0^r(T_A+T_B)G_0^r ~,\\
G_{0_{(j,j')}}^r &=& \frac{e^{ik|j-j'|}}{i\hbar v_k} \label{eq:g0} ~,\\
T_{A,B} &=& \frac{V_{A,B}}{1-G^r_0 V_{A,B}}~.
\end{eqnarray}
$T$ includes the influence of the potentials $V_{A,B}$ exactly. When
the potentials $V$ is small, then $T \approx V$ and $G$ is
approximately given as:
\begin{eqnarray}
 G^r &=& G_0^r +
G_0^r(V_A+V_B)G_0^r ~.\label{eq:gapp}
\end{eqnarray}
Inserting Eq. (\ref{eq:gapp}) into Eq. (\ref{eq:Jeq}) and retaining
the lowest order terms in $V$ leads to a useful form of $J$:
\begin{eqnarray}
J &=& \frac{1}{\pi}{\rm Im}~ {\rm Tr}\int
\left(V_B^\uparrow-V_B^\downarrow\right)G_0\left(V_A^\uparrow-V_A
^\downarrow\right) G_0 dE~.\label{eq:Jneq1} \nonumber \\
\end{eqnarray}

\begin{figure}[h]
\begin{center}
\vskip 0.2 cm
\includegraphics[width=3.25in]{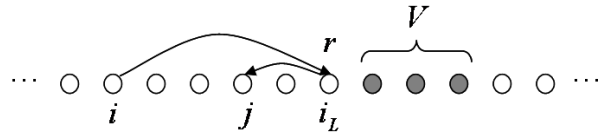}
\vskip 0.2 cm \caption{Schematic for effect of potential V on
$G_{i,j}$: the path from site i to site j can contain a reflection
$r$ off of the left edge of the potential V.}\label{fig:path}
\end{center}
\end{figure}

In the spirit of previous works on interlayer exchange coupling, we
next rewrite Eq. (\ref{eq:Jneq1}) in terms of spin-dependent
reflectivities of the magnetic layers. The effect of a single
potential $V_{A,B}$ is described by the reflection and transmission
coefficients $r_{A,B}$ and $t_{A,B}$. We suppose the potential to be
confined to sites $i$ within a range $i \in \{i'\}$, and that the
``left-most" site is $i_L$. For our purposes, it suffices to
consider the change in Green's function $\delta G_{i,j}$ induced by
the potential for indices $i,j$ on the same ``side" of the
potential, i.e. $i,j < i_L$. Then, for small $V_A$ (see Fig.
(\ref{fig:path})):
\begin{eqnarray}
\delta G_{i,j} &=& \left(G_0 T_A G_0 \right)_{i,j} \nonumber \\
&=&   i  \hbar v_k G_{0_{(i,i_L)}} r_A ~ G_{0_{(i_L,
j)}}~.\label{eq:Gfromr}
\end{eqnarray}

Using the fact that $V_{A,B} \approx T_{A,B}$, applying Eq.
(\ref{eq:Gfromr}) twice, and the form for $G_0$ given in Eq.
(\ref{eq:g0}) leads to:
\begin{eqnarray}
 J&=& \frac{1}{\pi}{\rm Im}\int \left(r^{\uparrow}_{A}-r^{\downarrow}_{A} \right)
\left(r^{\uparrow}_{B}-r^{\downarrow}_{B}\right) e^{2ik_{\rm F} D}
~dE~.\nonumber\\
\end{eqnarray}
where $D=a\left(n_{\rm S} + 1\right)$.  For the nonequilibrium case,
we insert Eq. (\ref{eq:gapp}) into Eq. (\ref{eq:jneq}), and keep
lowest order in $V$:
\begin{eqnarray}
J_{{\rm L,R}} &=& \frac{1}{\pi} {\rm Im}\int_{-\infty}^{\mu_{L,R}}
\left(V^\uparrow_B-V^\downarrow_B\right) \left( G_0^r \Gamma_{L,R}
G_0^a
\left(V^\uparrow_A-V^\downarrow_A\right) G_0^a \right. \nonumber \\
&&~~~~~\left.+ G_0^r \left(V^\uparrow_A-V^\downarrow_A\right) G_0^r
\Gamma_{L,R} G_0^a \right )dE~.
\end{eqnarray}
In terms of reflectivities and phase factors, this becomes:
\begin{eqnarray}
J_{\rm L} &=& 0 ~,\nonumber \\
J_{\rm R} &=& \frac{1}{\pi} \textrm{Im}\int_{- \infty}^{\mu_R}
\left[\left(r^{\uparrow}_{A}-r^{\downarrow}_{A} \right)
\left(r^{\uparrow}_{B}-r^{\downarrow}_{B}\right) e^{2ikD} \right]
dE~. \nonumber\\
\end{eqnarray}
Evaluating the integral in the large $D$ limit leads to Eq.
(\ref{eq:JsmallR}) of the text.


\begin{thebibliography}{00}
\bibitem{symmetryfootnote} In equilibrium, a spin valve with layer magnetizations in the
$\hat x-\hat z$ plane is invariant under the combined operations of
time reversal + spin rotation by $\pi$ about the $\hat y$ axis.
Out-of-plane torques are also invariant under this operation and are
therefore allowed by symmetry, while in-plane torques are not, and
therefore vanish.  The application of a current breaks time-reversal
symmetry, allowing a non-zero in-plane torque.  Similar arguments
can be made for general magnetization textures in equilibrium and
non-equilibrium.
\end{thebibliography}
\end{document}